\documentclass[slac_one]{revtex4}
\usepackage{graphicx}
\usepackage{fancyhdr}
\begin{document}

\title{Searches for hyperbolic extra dimensions at the LHC} 

%

\author{Tommy Ohlsson}
\affiliation{Department of Theoretical Physics, Royal Institute of Technology (KTH), 106 91 Stockholm, Sweden \&\\ Royal Swedish Academy of Sciences (KVA), P.O.~Box 50005, 104 05 Stockholm, Sweden}

\begin{abstract}
In this poster, we present a model of large extra dimensions where
the internal space has the geometry of a hyperbolic disc. Compared
with the ADD model, this model provides a more satisfactory solution
to the hierarchy problem between the electroweak scale and the Planck
scale, and it also avoids constraints from astrophysics. Since there
is no known analytic form of the Kaluza--Klein spectrum for our choice
of geometry, we obtain a spectrum based on a combination of
approximations and numerical computations. We study the possible
signatures of our model for hadron colliders, especially the LHC,
where the most important processes are the production of a graviton
together with a hadronic jet or a photon. We find that for the case of
hadronic jet production, it is possible to obtain relatively strong
signals, while for the case of photon production, this is much more
difficult.
\end{abstract}

\maketitle

\thispagestyle{fancy}


\section{INTRODUCTION AND THE HIERARCHY PROBLEM}

One of the main motivations for large extra dimensions is that they
provide a solution to the so-called hierarchy problem between the
electroweak scale $M_{\rm ew} \simeq 100 \, {\rm GeV}$ and the Planck
scale $M_{\rm Pl} \equiv G_N^{-1/2} \simeq 10^{19} \, {\rm
GeV}$. Theoretically, $M_{\rm ew}$ is expected to obtain loop
corrections of order $M_{\rm Pl} \gg M_{\rm ew}$. Therefore, a
miraculous cancellation is needed to keep $M_{\rm ew}$ at the order of
$100 \, {\rm GeV}$. The so-called ADD (Arkani-Hamed, Dimopoulos, and
Dvali) model \cite{arka98} provides a very elegant solution to this
problem. However, there is a problem related to this solution in the
ADD model, since it turns out that it just becomes a reformulation,
which means that the ADD model still suffers from the hierarchy
problem. Nevertheless, in models with hyperbolic geometry
\cite{kalo00}, this problem can be remedied.

\section{THE HYPERBOLIC DISC MODEL}

\begin{figure}[h]
\centering
\includegraphics[width=0.2\textwidth]{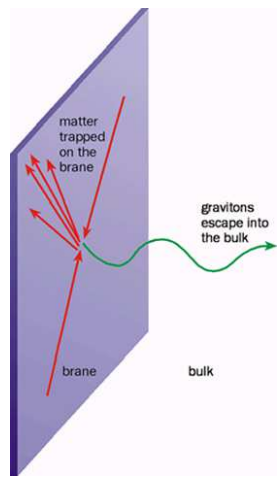}
\caption{An illustration of the brane and the extra dimensions. The
figure has been adopted with permission from \cite{iop00}.}
\label{fig:extradim}
\end{figure}

The model that we consider is similar to the ADD model, with the only
exception that the internal space is a two-dimensional hyperbolic
disc, with constant negative curvature $v$ \cite{melb08}. The SM
fields are assumed to be confined to a four-dimensional brane, while
gravity alone probes the extra dimensions. See Fig.~\ref{fig:extradim}
for an illustration.

In the ADD model, the Planck scale $M_{\rm Pl}$ is replaced by a new
mass scale $M_*$. The two scales are related through the equation
$M_{\rm Pl}^2 = V M_*^4$. If $V$ is large enough, then $M_*$ could be
as low as $M_{\rm ew} \simeq 1 \, {\rm TeV}$, hence eliminating the
hierarchy problem between $M_{\rm Pl}$ and $M_{\rm ew}$. However, in
the ADD model, the radius $L \simeq 10^{31/d} M_*$, and thus, a new
hierarchy problem is created, which means that the hierarchy problem
is only reformulated as the question of why the radius of the internal
space is so large compared to the electroweak scale

The area of a hyperbolic disc with radius $L$ is
\begin{displaymath}
V = \frac{4\pi}{v^2} \sinh^2 \left( \frac{vL}{2} \right) \sim
\frac{\pi}{v^2} \exp \left( vL \right).
\end{displaymath}
This has the result that, if the internal space is a hyperbolic disc,
then it is possible to have $M_{\rm Pl} \simeq M_{\rm ew} \simeq
L^{-1}$, {\it i.e.}, no new large hierarchy is introduced, if also $v
\simeq M_*$.

\section{COMPARISON WITH THE ADD MODEL}

Now, we perform a comparison between the ADD model and the hyperbolic
disc model. First, there are the properties of the ADD model: one free
parameter, {\it i.e.}, $M_*$; the Kaluza--Klein (KK) spectrum starts
out at $m \approx 0$; for $d=2$, $M_*$ is constrained to $M_* \geq 50
\, {\rm TeV}$ from astrophysics; universal coupling of KK modes;
physical results are independent of the position of the brane; and an
exact solution for the KK spectrum.  Second, there are the properties
of the hyperbolic disc model: three free parameters, {\it i.e.},
$M_*$, $v$, and $\tau_b$, where $\tau_b$ is the radial position of the
brane in the internal space; the KK spectrum starts out at $m \approx
v/2 > 0$; weak restrictions on the parameter space; different KK modes
have different couplings to SM fields; physical results depend on the
position of the brane; and no exact solution for the KK spectrum,
which means that approximations are needed. Thus, we observe that in
the hyperbolic disc model we have more parameter freedom, a mass gap,
and especially, the hierarchy problem can be solved. However, the
results will depend on the position of the brane and there is no
analytic solution for the KK spectrum.

\section{LHC PHENOMENOLOGY}

\begin{figure}[h]
\centering
\includegraphics[width=0.4\textwidth]{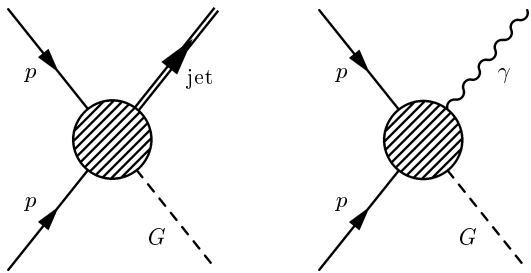}
\caption{The Feynman diagrams for the reactions $p + p \to {\rm jet} +
G$ (left) and $p + p \to \gamma + G$ (right).}
\label{fig:HEDfeynman}
\end{figure}

The KK modes of the graviton couple to all SM fields. We have
considered possible production of KK gravitons at the LHC. Since each
individual KK mode couples very weakly to SM fields, we need to study
the production of any kinematically available KK mode. Because of the
weak couplings, the graviton is not observed in detectors, and hence,
we consider production of a graviton together with some observable
particle. At the LHC, there are two interesting reactions: $p + p \to
{\rm jet} + G$ and $p + p \to \gamma + G$.  See
Fig.~\ref{fig:HEDfeynman} for the Feynman diagrams of the two
reactions.

\section{RESULTS}

For both of the reactions mentioned above, we have computed the
differential cross sections with respect to the transverse momentum
$p_{\rm T}$ of the outgoing jet/photon, and with respect to
$\cos(\theta)$, where $\theta$ is the angle between the beam and the
jet/photon. As a reference, we also give the corresponding results for
the ADD model, as well as the SM background, which mainly comes from
the processes $p + p \to {\rm jet}/\gamma + Z(\to \nu \bar \nu)$.

\begin{figure}[h]
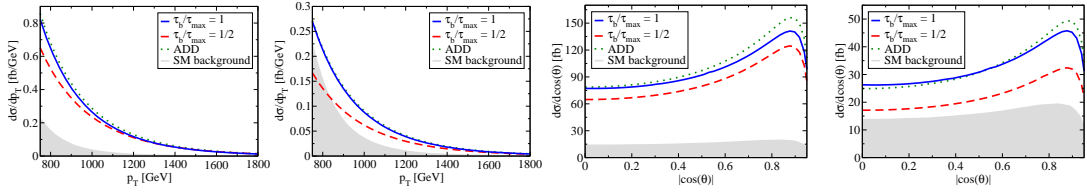

\centering
\includegraphics[width=0.4\textwidth,clip]{PresJetPT.eps}
\includegraphics[width=0.4\textwidth,clip]{PresJetCos.eps}
\caption{The differential cross section for graviton plus jet
production with respect to $p_{\rm T}$ and $\cos (\theta)$,
respectively. Left-left panel: $M_* = 1.5 \, {\rm TeV}$. Left-right
panel: $M_* = 2 \, {\rm TeV}$. Right-left panel: $M_* = 1.5 \, {\rm
TeV}$. Right-right panel: $M_* = 2 \, {\rm TeV}$.}
\label{fig:PresJet}
\end{figure}

\begin{figure}[h]
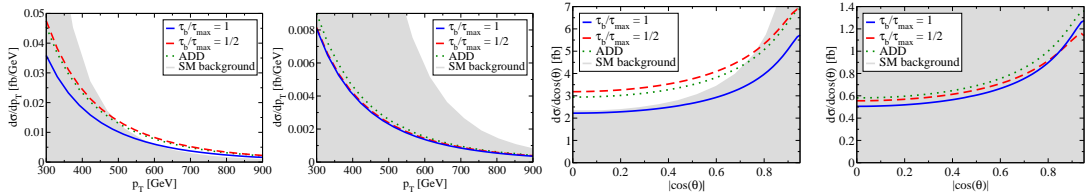

\centering
\includegraphics[width=0.4\textwidth,clip]{PresPhotonPT.eps}
\includegraphics[width=0.4\textwidth,clip]{PresPhotonCos.eps}
\caption{The differential cross section for graviton plus photon
production with respect to $p_{\rm T}$ and $\cos (\theta)$,
respectively. Left-left panel: $M_* = 1 \, {\rm TeV}$. Left-right
panel: $M_* = 1.5 \, {\rm TeV}$. Right-left panel: $M_* = 1 \, {\rm
TeV}$. Right-right panel: $M_* = 1.5 \, {\rm TeV}$.}
\label{fig:PresPhoton}
\end{figure}

In Fig.~\ref{fig:PresJet}, the jet production results that could be
observed at the LHC are shown as functions of $p_{\rm T}$ and
$\cos(\theta)$, respectively,
whereas in Fig.~\ref{fig:PresPhoton}, the photon production results
are shown as functions of $p_{\rm T}$ and $\cos(\theta)$,
respectively.

\section{CONCLUSIONS}

In conclusion, we have found that i) the signals are similar to those
of the ADD model, ii) the most promising signal comes from the jet
channel, and finally, iii) it is more difficult to obtain an
observable signal from the photon channel.

\begin{acknowledgments}
I would like to thank Henrik Melb{\'e}us for useful collaboration that
led to the publication \cite{melb08} upon which this poster is
based. In addition, I would like to thank the organizers of ICHEP 2008
for the possibility to show the poster.

This work was supported by the Royal Swedish Academy of Sciences (KVA)
and the Swedish Research Council (Vetenskapsr{\aa}det), Contract
No. 621-2005-3588.
\end{acknowledgments}

\end{document}